\documentclass[reprint,aps,prl,amsmath,showpacs]{revtex4-1}
\usepackage{graphicx}
\usepackage[linkcolor=blue,citecolor=blue]{hyperref}
\usepackage{bm}
\usepackage{float}
\usepackage{amssymb}
\usepackage{latexsym}
\hypersetup{colorlinks=true}

\def \qqquad {\qquad\quad}
\def \qqqquad {\qquad\qquad}

\newcommand{\mm}{\mu\bar{\mu}}

\begin{document}
\title{Hyperfine Splitting in True Muonium to $\mathcal{O}(m_\mu\alpha^6)$: Two Photon Annihilation Contribution from Other Flavors}

\author{Yao Ji}
\email{yao.ji@physik.uni-regensburg.de}
\affiliation{Institut f\"ur Theoretische Physik, Universit\"at Regensburg, Regensburg 93040, Germany}
\author{Henry Lamm}
\email{hlammiv@asu.edu}
\affiliation{Department of Physics, Arizona State University, Tempe, AZ 85287-1504, USA}
\date{\today}

\begin{abstract}
The two-photon-annihilation contribution to the true muonium hyperfine splitting arising from $e$ and $\tau$ loops is obtained analytically at order $m_\mu\alpha^6$.  The contribution to the hyperfine splitting is $-2.031092873 m_\mu\alpha^6/n^3\pi^2=-793.926988/n^3$ MHz.  The contribution to the triplet true muonium decay rate has also been obtained and was found to be $9.825708266 m_\mu\alpha^6/n^3\pi^2=3840.737698/n^3$ MHz.  Additional results have been computed for other purely leptonic bound states.
\end{abstract}

\maketitle
True muonium is the yet unidentified $(\mm)$ bound state.  This bound state has a metastable spectrum with lifetimes in the range of ps to ns~\cite{Brodsky:2009gx} since the 2.2 $\mu$s weak decay of the muon is much longer.  QED effects dominate the spectrum and transitions because the leptonic nature of true muonium suppresses QCD to vacuum polarization effects at $\mathcal{O}(m_\mu\alpha^5)$~\cite{Jentschura:1997tv,Jentschura:1997ma}.  Electroweak effects are suppressed further to $\mathcal{O}(m_\mu\alpha^7)$~\cite{PhysRevD.91.073008}.  The existing discrepancies in muon physics~\cite{PhysRevD.73.072003,Antognini:1900ns,Aaij:2014ora,Aaij:2015yra,Pohl669} motivate a serious investigation of true muonium, which can strongly discriminate between new physics models~\cite{TuckerSmith:2010ra,PhysRevD.91.073008,Lamm:2015fia,Lamm:2015gka}.  In the future, measurements of Lamb shift, $1s-2s$ splitting, and most relevant here the hyperfine splitting (hfs) should occur.  For competitive constraints from these experiments, Standard Model predictions are needed at the level of 100 MHz, corresponding to $\mathcal{O}(m_\mu\alpha^7)$.

The theoretical expression for the hfs corrections to true muonium from QED can be written
\begin{align}
 \Delta E_{\rm hfs}=m_\mu\alpha^4\bigg[&C_0+C_1\frac{\alpha}{\pi}+C_{21}\alpha^2\ln\left(\frac{1}{\alpha}\right)+C_{20}\left(\frac{\alpha}{\pi}\right)^2\nonumber\\
 &+C_{32}\frac{\alpha^3}{\pi}\ln^2\left(\frac{1}{\alpha}\right)+C_{31}\frac{\alpha^3}{\pi}\ln\left(\frac{1}{\alpha}\right)\nonumber\\
 &+C_{30}\left(\frac{\alpha}{\pi}\right)^3+\cdots\bigg],
\end{align}
where $C_{ij}$ indicate the coefficient of the term proportional to $(\alpha)^i\ln^j(1/\alpha)$.  All dependence of the hfs to mass scales other than $m_\mu$ is included in the $C_{ij}$.  The coefficients of single flavor QED bound states, used in positronium, are fully known up to $\mathcal{O}(m_e\alpha^6)$ and some partial results for $\mathcal{O}(m_e\alpha^7)$.  The leading-order $C_0=7/12$ was computed separately by Pirenne~\cite{Pirenne:1947}, Berestetskii~\cite{Berestetskii:1949}, and Ferrell~\cite{Ferrell:1951zz}.  The first order correction $C_1=-(1/2)\ln2-8/9$ was obtained by Karplus and Klein from loop corrections to the leading order calculations and the two-photon annihilation interaction~\cite{Karplus:1952wp}.  Conceptual and mathematical difficulties arose in deriving the first logarithmic coefficient, but the result, $C_{21}=5/24$, was eventually obtained by Lepage~\cite{Lepage:1977gd}.  The purely $m_e\alpha^6$ dependent coefficient took more than two decades to complete due to the shear number of contributions, but was found to be $C_{20}=-\frac{52}{32}\zeta(3)+\left(\frac{221}{24}\ln(2)-\frac{5197}{576}\right)\zeta(2)+\frac{1}{2}\ln(2)+\frac{1367}{648}$ (see \cite{Czarnecki:1998zv} and the references therein).  

At $\mathcal{O}(m_e\alpha^7)$, $C_{32}=-7/8$ was found by Karshenboim~\cite{Karshenboim:1993} in 1993, and $C_{31}=-17/3\ln2+217/90$ was found by several groups in 2000~\cite{Kniehl:2000cx,Melnikov:2000zz,Hill:2000zy}.  Today, only partial results for the $m_e\alpha^7$ coefficient exist~\cite{marcu2011ultrasoft,Baker:2014sua,Adkins:2014dva,Eides:2014nga,Eides:2015nla,Adkins:2014xoa,Adkins:2015jia,Adkins:2015dna,Adkins:2016btu}, which total to $C_{30,{\rm partial}}\approx160$ at present.  These results can be translated to true muonium with the exchange $m_e\rightarrow m_\mu$.

In addition to these, true muonium has extra contributions that must be considered.  The existence of the lighter electron allows for large loop contributions to true muonium system.  The relative smallness of $m_\tau/m_\mu\approx 17$ and $m_{\pi}/m_{\mu}\approx1.3$ produce contributions to true muonium much larger than analogous contributions to positronium.  Of these true muonium specific contributions, which we denote by $C_{ij}^\mu$, only a few terms are known.  $C_{1}^\mu=1.638(5)$ was computed by Jentschura et. al.~\cite{Jentschura:1997tv} where only the electron and hadronic loops have been considered since the tau contributions are numerically smaller than the uncertainty, which is estimated from the model dependence of the hadronic contribution.  Until this Letter, no computed $\mathcal{O}(m_\mu\alpha^6)$ contributions existed, but the electron loop in three-photon annihilation at $\mathcal{O}(m_\mu\alpha^7)$ is $C_{30,3\gamma}^\mu=-5.86510(20)$~\cite{Adkins:2015jia}.  For a $\mathcal{O}(m_\mu\alpha^7)$ prediction of the hfs, contributions from $Z$-bosons must be considered because of its $m_\ell^3$ scaling~\cite{PhysRevD.91.073008}, allowing for a low energy determination of $\sin\theta_W$.

\begin{figure}
\includegraphics{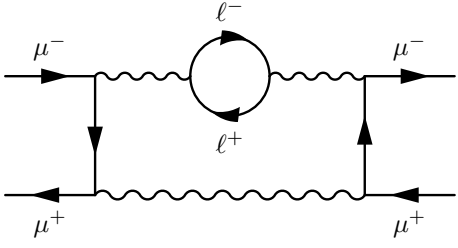}
\caption{\label{fig:2gloop}A example of a general leptonic loop in the two-photon annihilation graph of true muonium.}
\end{figure}

In this Letter, we compute the first piece of $C_{20}^\mu$, the two-photon annihilation contributions arising from a single lepton loop (see Fig.~\ref{fig:2gloop}).  We neglect the hadronic contribution since the model-dependent nature of present techniques must be improved already\cite{Jentschura:1997tv}.   We work initially with a general $\beta=(m_\ell/m_{\ell'})^2$, and at the end consider the cases of physical relevance $\ell,\ell'=e,\mu,\tau$.  This calculation generalizes the results of~\cite{Adkins:1993zz} where the two-photon annihilation correction was considered for $\beta=1$ in the Fried-Yennie gauge~\cite{Fried:1958zz}. The effect of vacuum polarization on the photon propagator is given by $\frac{1}{k^2}\rightarrow -\frac{1}{k^2}\Pi_R(k^2)$, where the renormalized $\mathcal{O}(\alpha)$ vacuum polarization factor is
\begin{equation}
 \Pi_R(k^2)=-\frac{\alpha}{3\pi}\frac{k^2}{m^2}\int^1_0\mathrm{d}x\frac{x^2(3-8x+4x^2)}{1-x(1-x)k^2/m^2}.
\end{equation}
Using this, the full correction to energy can be found to be\cite{Adkins:1993zz,Adkins:2016btu}
\begin{align}
 \Delta E_{\rm VP}&=\frac{m_\mu\alpha^5}{n^3\pi}(2)\int\frac{\mathrm{d}k^4}{i\pi^2}\frac{-\bm{k}^2(-1)\Pi_{R}(m^2k^2)}{k^2(k-2N)^2(k^2-2kN)^2}\nonumber\\
 &=\frac{m_\mu\alpha^6}{n^3\pi^2}I_{\rm VP},
\end{align}
where the four-vectors $k$ and $N=(1,\bm{0})$ are normalized to 1 by scaling out the mass $m$ of the muons and the factor of 2 arises from the possibility of inserting the vacuum polarization bubble into either photon line.  Comparing to Eq.~(26) of~\cite{Adkins:1993zz}, the general $\beta$ case for $I_{\rm VP}$ is   
\begin{align}
 I_{\rm VP}=-\frac{2}{3}&\int^1_0\mathrm{d}x\int\frac{\mathrm{d}k^4}{i\pi^2}\frac{x(3-8x+4x^2)}{(1-x)}\nonumber\\\times&\frac{-\bm{k}^2}{k^2(k^2-2kN)^2\left(-\frac{1}{\beta x(1-x)}+ (2N-k)^2\right)}.
 \end{align}
Performing a standard Feynman parameterization allows for integration without issue over $k^2$.  The resulting integral is completely finite and found to be
 \begin{equation}
 \label{eq:integral}
 I_{\rm VP}=\int^1_0\mathrm{d}x\mathrm{d}y\int^{1-y}_0dz\,\frac{\beta x^2y(3-8x+4x^2)}{z+\beta x(1-x)\big[(y+2z)^2-4z\big]}.
\end{equation}

 In solving the integral in Eq.~(\ref{eq:integral}), the most challenging part we encountered involves integral of the form,
\begin{align}
 I_c&=-\frac{1}{24\beta}\bigg\{\int^{1}_0dx\,\frac{4\text{arctanh}\Big(\sqrt{1-4\beta x(1-x)}\Big)}{(1-x)^2\sqrt{1-4\beta x(1-x)}}\nonumber
\\
&+\int^{1}_0\frac{dx}{(1-x)^2}\,\bigg\{2\Big[2\beta x(1-x)\Big(2+\ln\big(\beta x(1-x)\big)\Big)\nonumber
\\
&\qquad+\ln\big(\beta x(1-x)\big)\Big]+\big(1-4\beta x(1-x)\big)\nonumber\\
&\qqqquad\qqqquad\qqqquad\times\ln\big(1-4\beta x(1-x)\big)\bigg\}\nonumber
\\
&=-\frac1{24\beta}\bigg\{I_{c1}+I_{c2}\bigg\}\, .
\end{align}

The integral $I_c$ is finite, but to simplify the computation, it is convenient to calculate $I_{c1}$ and $I_{c2}$ separately.  This however is complicated by $I_{c1}$ and $I_{c2}$ being separately divergent. In order to control the divergences, we employ the following limitation approach,
 \begin{align}
 I_c&=-\frac{1}{24\beta}\lim_{\epsilon\rightarrow0^+}\bigg\{\int^{1-\epsilon}_0dx\,\frac{4\text{arctanh}\Big(\sqrt{1-4\beta x(1-x)}\Big)}{(1-x)^2\sqrt{1-4\beta x(1-x)}}\nonumber
\\
&+\int^{1-\epsilon}_0\frac{dx}{(1-x)^2}\,\bigg\{2\Big[2\beta x(1-x)\Big(2+\ln\big(\beta x(1-x)\big)\Big)\nonumber
\\
&\qquad+\ln\big(\beta x(1-x)\big)\Big]+\big(1-4\beta x(1-x)\big)\nonumber\\
&\qqqquad\qqqquad\qqqquad\times\ln\big(1-4\beta x(1-x)\big)\bigg\}\nonumber
\\
&=-\frac1{24\beta}\bigg\{I^\epsilon_{c1}+I^\epsilon_{c2}\bigg\}\, .
 \end{align}
where $\epsilon\rightarrow0^+$ means that the limitation is carried out from above. With this procedure, the $I_{c2}^\epsilon$ integral can be computed in a straightforward fashion and we thusly focus on $I_{c1}$ in the following.

Defining new variables $a=\sqrt{1-\beta}$ and $b=\displaystyle\sqrt{\frac\beta{1-\beta}}$ then we have via change of variables,
\begin{widetext}
\begin{align}
\label{eq:Ic1}
I_{c1}^\epsilon&=\frac1{3\sqrt\beta(1-\beta)}\int^{\tan\big(\frac12\text{arccot}\big(\frac{\sqrt{1-\beta}}{\sqrt{\beta}}\big)-\frac{\epsilon_\alpha}2\big)}_{-\tan\big(\frac12\text{arccot}\big(\frac{\sqrt{1-\beta}}{\sqrt{\beta}}\big)\big)}dt\,\frac{(1-t^2)\big[\ln(1-a-(1+a)t^2)-\ln(1+a-(1-a)t^2)\big]}{\big(2t-b(1-t^2)\big)^2}\, ,
\end{align}
\end{widetext}
where $\displaystyle\frac12-\epsilon= \displaystyle\sqrt{\frac{1-\beta}{4\beta}}\tan\big(\text{arccot}\big({\sqrt{1-\beta}}/{\sqrt\beta}\big)-\epsilon_\alpha\big)$.

To compute the indefinite integral with integrand proportional to $\ln(1-a-(1+a)t^2)$ shown above, we employ the following trick,
\begin{align}
\label{eq:indefiniteint}
\tilde I_{c_1}^\epsilon&=\int dt\,\frac{(1-t^2)\ln(1-a-(1+a)t^2)
}
{\big(2t-b(1-t^2)\big)^2
}\,\nonumber
\\
&=\frac\partial{\partial b}\frac1{b(b_2-b_1)}\int dt\,\Big(\frac1{t-b_2}-\frac1{t-b_1}\Big)\Big[\ln(1+a)\nonumber
\\
&\qqquad+\ln(t-a_1)+\ln(t-a_2)-\ln(1-a)\Big]\, ,
\end{align}
where $\{a_1\,,a_2\}$ 
and $\{b_1\,,b_2\}$ are solutions to algebraic equations $(1+a)t^2+a-1=0$ and $2t-b(1-t^2)=0$, respectively. 
Then the integration of Eq.~(\ref{eq:indefiniteint}) becomes elementary. The contribution involving $\ln(1+a-(1-a)t^2)$ in Eq.~(\ref{eq:Ic1}) is obtained from Eq.~(\ref{eq:indefiniteint}) by replacing $a$ with $-a$. It is worth noting that the antiderivative obtained directly from Eq.~(\ref{eq:indefiniteint}) is divergent when $b^2=\displaystyle\frac{1-a^2}{a^2}$, which is the case we are concerned with. To deal with this issue, we write $b^2=\displaystyle\frac{1-a^2}{a^2}+\epsilon$ and pull out the divergent constant piece proportional to $1/\epsilon$ which approaches infinity when $b^2=\displaystyle\frac{1-a^2}{a^2}$. 

Putting all of the pieces together, the final result of $I_{\text{VP}}$ reads, 
\begin{widetext}
\begin{align}
\label{eq:energy}
  I_{\rm VP}&= \frac{1}{72\beta} \bigg\{\frac{3 \ln\beta}{\sqrt{\beta }} \bigg[\left(1-2 \beta ^{3/2}\right) \ln (1-\beta )-2 \ln \big((1-\sqrt{1-\beta })
  (\sqrt{\beta }+1)\big)+6 \ln \big(\sqrt{1-\beta }+\sqrt{\beta }-1\big)+\ln\beta\nonumber
  \\
  &\quad-4 \text{arctanh}\left(\sqrt{1-\beta
   }-\sqrt{\beta }\right)\bigg]+3 \ln (1-\beta ) \bigg[2 \beta  \ln \left(\frac{1-\sqrt{1-\beta }}{\sqrt{1-\beta }}\right)+\frac{4 \ln
   \left(\sqrt{\beta }+1\right)}{\sqrt{\beta }}-\frac{2 \ln \left(\sqrt{1-\beta }+\sqrt{\beta }-1\right)}{\sqrt{\beta }}\nonumber
  \\
  &\quad+\left(\beta
   -\frac{1}{\sqrt{\beta }}\right) \ln (1-\beta )\bigg]+8 (\beta +3) \ln\beta+6 \beta  \text{arccosh}(1-2 \beta )^2+\frac{24 \ln
   \left(\sqrt{1-\beta }+\sqrt{\beta }-1\right) \text{arctanh}\left(\sqrt{1-\beta }-\sqrt{\beta }\right)}{\sqrt{\beta }}\nonumber
   \\
   &\quad-\frac4{\sqrt\beta} \bigg[\pi ^2
   \left(\beta^{3/2} -2\right)-16 \sqrt\beta(\beta -1)+3 \ln ^2\left(\sqrt{\beta }+1\right)-3 \ln \left(\sqrt{1-\beta
   }+\sqrt{\beta }-1\right) \ln \left(\frac{\sqrt{\beta }+1}{\beta(1+\sqrt{1-\beta })}\right)\nonumber
   \\
   &\quad+4 \sqrt{1-\beta } \left(3
   \beta  \text{arcsin}\sqrt{\beta }+(\beta -1) \text{arccot}\left(\frac{\sqrt{1-\beta }}{\sqrt{\beta }}\right)\right)\bigg]+12 \bigg[\beta  \bigg(\text{Li}_2\left(\frac{1}{\sqrt{1-\beta }+1}\right)+2 \text{Li}_2\left(\sqrt{1-\beta
   }\right)\nonumber
   \\
   &\quad-\text{Li}_2\left(\frac{\beta +\sqrt{1-\beta }-1}{\beta }\right)\bigg)-\frac{2}{\sqrt{\beta }} \Big[2 \text{Li}_2\left(\frac{\sqrt{1-\beta
   }}{\sqrt{\beta }+1}\right)+\text{Li}_2\left(\frac{1-\sqrt{1-\beta }}{\sqrt{\beta }}\right)-\text{Li}_2\left(\frac{\sqrt{1-\beta }-1}{\sqrt{\beta
   }}\right)\nonumber
   \\
   &\quad-\text{Li}_2\left(\frac{\left(\sqrt{1-\beta }-1\right) \left(\sqrt{\beta }-1\right)}{\sqrt{(1-\beta ) \beta
   }}\right)+\text{Li}_2\left(\frac{\beta +\sqrt{1-\beta }-1}{\beta -\sqrt{\beta }}\right)\Big]\bigg]\bigg\}\,,
\end{align}
\end{widetext}

where the expression is real for $0 \leqslant\beta\leqslant1 $ whereas for the $\beta>1$ region, the expression picks up an imaginary part corresponding to on-shell decay. In order to obtain consistent results for $\beta>1$ with the pole prescription of the Feynman propagator and obtain the correct branch cut, we let $\beta\rightarrow\beta+i\epsilon$.  Taking the limit of $\beta\rightarrow 1$, we recover the result of~\cite{Adkins:1993zz}, $I_{VP}=-\frac{1}{6}\zeta(2)$.  

Two simplifying limits are useful to consider for the real parts of $\Delta E_{\rm VP}$.  The first is $\beta\rightarrow0$ where the leptons running in the loop become infinitely massive
\begin{equation}
 \Delta E_{{\rm VP},0}=\frac{m_\mu\alpha^6}{n^3\pi^2}\left[\frac{\beta(-31+15\ln\beta)}{150}+\mathcal{O}(\beta^2)\right],
\end{equation}
 and $\beta\rightarrow\infty$ where the leptons in the loop become nearly but not quite massless (allowing the leptons to become massless introduces an infrared divergence)
\begin{align}
\Delta E_{{\rm VP},\infty}=&\frac{m_\mu\alpha^6}{n^3\pi^2}\bigg[\frac{1}{36}\bigg(-5\pi^2+4(8-\ln2[8 -\ln8])\nonumber\\&+12(-1+\ln2)\ln\beta\bigg)+\mathcal{O}\left(\beta^{-1}\right)\bigg]
\end{align}%
where each agrees within 1\% with the exact solution for physical values of $\beta$.  To obtain the contributions to the triplet state decay rate, we use the definition $\Gamma=-2\text{Im}(\Delta E_{\rm VP})$.  Our results agrees with those found in~\cite{Jentschura:1997tv} where the asymptotic form of the vacuum polarization was used.  In Table~\ref{tab:values} we have listed the numerical values of $I_{\rm VP}$ for the physical values of $\beta$.
Using Eq.~(\ref{eq:energy}), we see that the energy shift in true muonium due to $e$ is $\Delta E_{{\rm VP},e}=\frac{-792.859944}{n^3}\text{~MHz}$,
and the contribution to the decay rate of $\Gamma_{{\rm VP},e}=9.825708266\frac{m_\mu\alpha^6}{n^3\pi^2}=\frac{3840.737698}{n^3}\text{~MHz}$.  The much smaller contribution from $\tau$ is found to be $\Delta E_{{\rm VP},\tau}=-\frac{1.067044}{n^3}\text{~MHz}.$  
\begin{table}
 \caption{\label{tab:values}Two-photon annihilation contributions from leptonic loops to the hfs, $\Delta E_{\rm VP}$ in units of $m_\mu\alpha^6/\pi^2$ for physical values of $\beta_{ij}$ where $i$ is the valence lepton, and $j$ is the lepton in the loop.}
 \begin{center}
 \begin{tabular}{l c}
 \hline\hline
  $\beta$&$I_{\rm VP}$\\
  \hline
  $\beta_{e\tau}=8.3\times10^{-8}$&$-1.519746398\times10^{-7}$\\
  $\beta_{e\mu}=2.3\times10^{-5}$&$-2.977551560\times10^{-5}$\\
  $\beta_{\mu \tau}=3.5\times10^{-3}$&$-2.729803616\times10^{-3}$\\
  $\beta_{\ell\ell }=1$&$-2.741556778\times10^{-1}$\\
  $\beta_{\tau \mu}=2.8\times10^2$&$-1.500492860-2.287953105i$\\
  $\beta_{\mu e}=4.3\times10^{4}$&$-2.028363069 -4.912854133 i$\\
  $\beta_{\tau e}=1.2\times10^{7}$&$-2.605913761 -7.868414309 i$\\
  \hline\hline
 \end{tabular}
\end{center}
\vspace{-0.9cm}
\end{table}
 With these contribution found, it is useful to reevaluate the uncertainty estimate of~\cite{PhysRevD.91.073008}.  The electron vacuum polarization corrections, which are the largest contribution at $\mathcal{O}(m_\mu\alpha^5)$\cite{Jentschura:1997tv,Jentschura:1997ma,Karshenboim:1998am},  were used to estimate the unknown $\mathcal{O}(m_\mu\alpha^6)$ corrections unique to true muonium by multiplying the complete $\mathcal{O}(m_\mu\alpha^5)$ diagrams by the photon polarization function, $\Pi(q^2)$, that arises from the electron vacuum polarization at momentum $q^2=4m_\mu^2$.  This method ignores the proper convolution and non-electronic contributions, but should give a gross estimate.  The majority of $\mathcal{O}(m_\mu\alpha^5)$ contain two photon propagators, so this estimate is multipled by 2,
\begin{align}
 \delta E^{6}_{\rm hfs}|_{\mu}\approx& 2\Pi_R(4m_\mu^2)\Delta E^{5}_{\rm hfs,remain}&\nonumber\\
 \approx&6.92\left(-\frac{25}{19}+1.638(5)\right)\frac{m_\mu\alpha^6}{\pi^2}\approx700 \text{MHz},
\end{align}
where $\Delta E^{5}_{\rm hfs}$ was obtained in \cite{Jentschura:1997tv} and here consists of the sum of the $C_{1}$ and $C_1^{\mu}$ terms except for the two-photon annihilation term.  Comparing to~\cite{PhysRevD.91.073008}, our error estimate is reduced by $500$ MHz.  With the correction now known to be $-764$ MHz, the uncertainty estimate was accurate to within a factor of two, encouraging us that this method is reasonable.  With the newly computed term and error estimate, we find $\Delta E^{1s}_{\rm hfs}=42329730(800)(700)\text{ MHz}$,
where the first uncertainty is from hadronic model dependence, and the second a revised estimate of missing $\mathcal{O}(m_\mu\alpha^6)$ terms.

HL would like to thank G. Adkins for his help with calculation issues.  HL is supported by the National Science Foundation under Grant Nos. 
PHY-1068286 and PHY-1403891.  YJ acknowledges the Deutsche Forschungsgemeinschaft for support under grant BR 2021/7-1.
\bibliography{/home/hlamm/wise}
\end{document}